\documentstyle{aipproc}
\input{psfig}
\def\Journal#1#2#3#4{{#1} {\bf #2}, #3 (#4)}

\def\NPA{{\em Nucl. Phys.} A}
\def\PLB{{\em Phys. Lett.}  B}
\def\PRL{\em Phys. Rev. Lett.}
\def\PRD{{\em Phys. Rev.} D}

\def\ZPC{{\em Z. Phys.} C}
\def\PAN{\em Phys. At. Nuclei }
\def\EPJ{{\em Eur. Phys. J.} C} 
\begin{document}
\title{Exploring Timelike Region of QCD Exclusive Processes in Relativistic
Quark Model}

\author{Ho-Meoyng Choi and Chueng-Ryong Ji}
\address{Department of Physics, North Carolina State University,
Raleigh, NC 27695-8202}

\maketitle

\begin{abstract}
We investigate the form factors and decay rates of exclusive 
$0^{-}\to0^{-}$ semileptonic meson decays using the constituent 
quark model based on the light-front quantization. 
Our model is constrained by the variational principle for the 
linear plus Coulomb interaction motivated by QCD. 
Our numerical results are in a good agreement with the available 
experimental data. 
\end{abstract}

%
%
%
%
One of the distinctive advantages in the light-front approach is the
well-established formulation of various form factor calculations
using the well-known Drell-Yan-West ($q^{+}$=0) frame\cite{LB}.
In $q^{+}$=0 frame, only parton-number-conserving Fock
state (valence) contribution is needed when the ``good" components of the
current, $J^{+}$ and $J_{\perp}$=$(J_{x},J_{y})$, are used\cite{Kaon}.
For example, only the valence diagram shown in Fig. 1(a) is used in the
light-front quark model (LFQM) analysis of spacelike meson form factors.
Successful LFQM description of various hadron form
factors can be found in the literatures\cite{CJC,NPA,Mix}.

However, the timelike ($q^{2}>0$) form factor analysis in the LFQM
has been hindered by the fact that $q^{+}$=0 frame is defined
only in the spacelike region ($q^{2}$=$q^{+}q^{-}-q^{2}_{\perp}<0$).
While the $q^{+}$$\neq$0 frame can be used in principle to compute the
timelike form factors, it is inevitable (if $q^{+}$$\neq$0) to encounter the
nonvalence diagram arising from the quark-antiquark pair creation (so called
``Z-graph"). For example, the nonvalence diagram in the case of
semileptonic meson decays is shown in Fig. 1(b).
The main source of the difficulty, however, in calculating the nonvalence
diagram(see Fig. 1(b)) is the lack of information on the black blob
which should contrast with the white blob representing the usual light-front
valence wave function. In fact, we noticed\cite{Kaon} that
the omission of nonvalence contribution
leads to a large deviation from the full results.
\begin{figure}[t]
\centerline{\psfig{figure=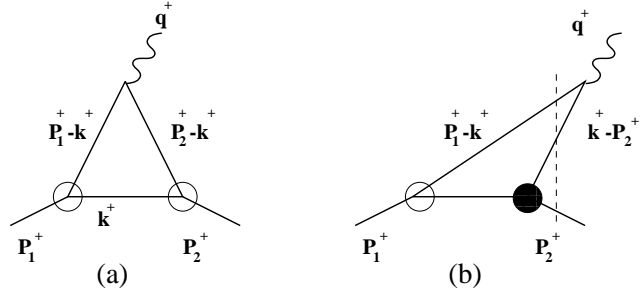,height=1.5in}}
\caption{The LFQM description of a electroweak
meson form factor: (a) the usual light-front valence diagram and (b) the
nonvalence(pair-creation) diagram. The vertical dashed line in (b)
indicates the energy-denominator for the nonvalence contributions.
While the white blob represents the usual light-front
valence wave function, the modeling of black blob has not yet been made.
\label{fig:triangle}}
\end{figure}

In this paper, we circumvent this problem by calculating the 
semileptonic processes in $q^{+}$=0 frame and then analytically  
continuing to the timelike region.
The $q^{+}$=0 frame is useful because only
valence contributions are needed. However, one needs to calculate
the component of the current other than $J^{+}$ to obtain the form factor
$f_{-}(q^{2})$. Since $J^{-}$ is not free from the zero-mode contributions
even in $q^{+}$=0 frame\cite{Zm,BH},
we use $J_{\perp}$ instead of $J^{-}$ to obtain $f_{-}$.

The key idea in our LFQM\cite{Mix} for mesons is to treat
the radial wave function as a trial function for the variational
principle to the QCD-motivated Hamiltonian saturating
the Fock state expansion by the constituent quark and
antiquark. The spin-orbit wave function is uniquely determined by the
Melosh transformation. We take the QCD-motivated effective Hamiltonian
as the well-known linear plus Coulomb interaction given by
\begin{eqnarray}
H_{q\bar{q}}= H_{0} + V_{q\bar{q}}
=\sqrt{m_{q}^{2}+k^{2}} + \sqrt{m_{\bar{q}}^{2}+k^{2}}+ V_{q\bar{q}},
\end{eqnarray}
where
\begin{eqnarray}
V_{q\bar{q}}= V_{0} + V_{\rm hyp}
= a + br - \frac{4\kappa}{3r}
+ \frac{2\vec{S}_{q}\cdot\vec{S}_{\bar{q}}}
{3m_{q}m_{\bar{q}}}\nabla^{2}V_{\rm Coul}.
\end{eqnarray}
We take the Gaussian radial wave function
$\phi(k^{2})=N\exp(-k^{2}/2\beta^{2})$
as our trial wave function to minimize the central Hamiltonian\cite{Mix}.
Since the string tension $b$=0.18 GeV$^{2}$ and the constituent $u$ and
$d$ quark masses $m_{u}$=$m_{d}$=0.22 GeV are rather well known from
other quark model analyses commensurate with Regge
phenomenology\cite{Isgur}, we take them as our input parameters.
The model parameters of $a,\kappa$, and $\beta_{u\bar{d}}$
are determined by the variational principle using the masses
of $\rho$ and $\pi$~\cite{Mix,PLB}. It is very important to note that
all other model parameters such as $m_{c}$, $m_{b}$,
$\beta_{uc}$, $\beta_{ub}$, etc. are then uniquely determined by
our variational principle as shown in \cite{PLB}. 
More detailed procedure of determining the model parameters and
ground state meson mass spectra can be found in \cite{Mix,PLB}.

Our predictions of the ground state meson mass spectra\cite{PLB}
are in a good agreement with the available experimental data. 
Furthermore, our model predicts the two unmeasured mass
spectra of $^{1}S_{0}(b\bar{b})$ and $^{3}S_{1}(b\bar{s})$ systems as
$M_{b\bar{b}}$=9657 MeV and $M_{b\bar{s}}$=5424 MeV, respectively.
Our values of the decay constants\cite{PLB} are also in a good
agreement with the results of lattice QCD\cite{Flynn}
anticipating future accurate experimental data.

The matrix element of the current $j^{\mu}=\bar{q}_{2}\gamma^{\mu}Q_{1}$
for $0^{-}(Q_{1}\bar{q})\to0^{-}(q_{2}\bar{q})$ decays  
can be parametrized in terms of two hadronic form factors as follows
\begin{eqnarray}
\langle P_{1}|\bar{q}_{2}\gamma^{\mu}Q_{1}|P_{1}\rangle&=&
f_{+}(q^{2})(P_{1}+P_{2})^{\mu} + f_{-}(q^{2})(P_{1}-P_{2})^{\mu},
\nonumber\\
&=& f_{+}(q^{2})\biggl[(P_{1}+P_{2})^{\mu}
-\frac{M^{2}_{1}-M^{2}_{2}}{q^{2}}q^{\mu}\biggr]
+ f_{0}(q^{2})\frac{M^{2}_{1}-M^{2}_{2}}{q^{2}}q^{\mu},
\end{eqnarray}
where $q^{\mu}=(P_{1}-P_{2})^{\mu}$ is the four-momentum
transfer to the leptons and $m^{2}_{l}\leq q^{2}\leq(M_{1}-M_{2})^{2}$.
The form factors $f_{+}$ and $f_{0}$ are related to the exchange of $1^{-}$
and $0^{+}$, respectively, and satisfy the following relations:
\begin{eqnarray}
f_{+}(0)&=& f_{0}(0),\hspace{.2cm}f_{0}(q^{2})= f_{+}(q^{2})
+ \frac{q^{2}}{M^{2}_{1}-M^{2}_{2}}f_{-}(q^{2}).
\end{eqnarray}
In the LFQM calculations presented in Ref.\cite{Dem}, the
$q^{+}$$\neq$0 frame has been used to calculate the semileptonic decays
in the timelike region.
However, when the $q^{+}$$\neq$0 frame is used, the inclusion of the
nonvalence contributions arising from quark-antiquark pair
creation (see Fig. 1(b)) is inevitable and this inclusion may be
very important for light-to-light and heavy-to-light decays.
Nevertheless, the previous analyses~\cite{Dem} considered only valence 
contributions in $q^{+}$$\neq$0 frame neglecting nonvalence
contributions. In this work, we circumvent this problem by calculating
the processes in $q^{+}$=0 frame and analytically continuing to the
timelike region. The $q^{+}$=0 frame is useful because only
valence contributions are needed. However, one needs to calculate
the component of the current other than $J^{+}$ to obtain the form factor
$f_{-}(q^{2})$. Since $J^{-}$ is not free from the zero-mode
contributions even in $q^{+}$=0 frame~\cite{Zm},
we use $J_{\perp}$ instead of $J^{-}$ to obtain $f_{-}$.
In the $q^{+}$=0 frame, we obtain the form factors $f_{+}(q^{2})$
and $f_{-}(q^{2})$ using the matrix element of the ``$+$" and
``$\perp$"-components of the current, $J^{\mu}$, respectively, and
then analytically continue to the timelike $q^{2}>0$ region by
changing $q_{\perp}$ to $iq_{\perp}$ in the form factors.

\paragraph*{Light-to-light decays:}

For $K_{l3}$ decays, the three form factor parameters,
i.e., $\lambda_{+}, \lambda_{0}$ and $\xi_{A}$,
have been measured using the following linear parametrization~\cite{data}:
\begin{eqnarray}
f_{\pm}(q^{2})&=& f_{\pm}(q^{2}=m^{2}_{l})\biggl(
1 + \lambda_{\pm}\frac{q^{2}}{M^{2}_{\pi^{+}}}\biggr),
\end{eqnarray}
where $\lambda_{\pm,0}$ is the slope of $f_{\pm,0}$ evaluated at
$q^{2}=m^{2}_{l}$ and $\xi_{A}=f_{-}/f_{+}|_{q^{2}=m^{2}_{l}}$.
\begin{table}
\caption{
Model predictions for the parameters of $K_{l3}$ decay form factors
obtained from $q^{+}$=0 frame. The charge radius $r_{\pi K}$ is obtained
by $\langle r^{2}\rangle_{\pi K}=6f'_{+}(q^{2}=0)/f_{+}(0)$.
For comparison, we include the results (in square brackets) of the valence 
contribution in $q^{+}$$\neq$0 frame. The CKM matrix used in the calculation 
of the decay width (in units of $10^{6}$ s$^{-1}$) is
$|V_{us}|=0.2205\pm0.0018$ [12].}
\begin{tabular}{cccr}
Observables & Our model & Other models & Experiment\\
\tableline
$f_{+}(0)$ & 0.962[0.962] & $0.961\pm0.008^{\cite{Roos}}$,
$0.952^{\cite{Buck}}$,$0.93^{\cite{SI}}$ & \\ 
$\lambda_{+}$& 0.026[0.083] & 0.028$^{\cite{Buck}}$,0.019$^{\cite{SI}}$
& $0.0286\pm0.0022[K^{+}_{e3}]$\\
& & & $0.0300\pm0.0016[K^{0}_{e3}]$\\   
$\lambda_{0}$& $-0.009[-0.017]$ &
0.0026$^{\cite{Buck}}$,$-0.005^{\cite{SI}}$ &
$0.004\pm0.007[K^{+}_{\mu3}]$\\
& & & $0.025\pm0.006[K^{0}_{\mu3}]$\\  
$\xi_{A}$& $-0.41[-1.10]$ &
$-0.28^{\cite{Buck}}$,$-0.28^{\cite{Buck}}$ &
$-0.35\pm0.15[K^{+}_{\mu3}]$\\
& & & $-0.11\pm0.09[K^{0}_{\mu3}]$\\  
$\langle r\rangle_{\pi K}$(fm) & 0.56 &
0.57$^{\cite{Buck}}$, 0.48$^{\cite{SI}}$ & \\  
$\Gamma(K^{0}_{e3})$ &
$7.36\pm 0.12$ & &$7.7\pm 0.5[K^{0}_{e3}]$\\
\end{tabular}
\end{table}
Our predictions of the parameters for $K_{l3}$ decays in $q^{+}$=0
frame, i.e., $f_{+}(0)$, $\lambda_{+}$, $\lambda_{0}$,
$\langle r^{2}\rangle_{K\pi}=6f'_{+}(0)/f_{+}(0)= 
6\lambda_{+}/M^{2}_{\pi^{+}}$, and
$\xi_{A}=f_{-}/f_{+}|_{q^{2}=m^{2}_{l}}$,
are summarized in Table 1.
Our result for the form factor $f_{+}$ at zero momentum
transfer, $f_{+}(0)=0.962$, is consistent with
the Ademollo-Gatto theorem\cite{Gatto} and also in a good agreement
with the result of chiral perturbation theory\cite{Roos},
$f_{+}(0)=0.961\pm 0.008$. Our results for other observables such as
$\lambda_{+}$, $\xi_{A}$, and $\Gamma(K_{l3})$
are overall in a good agreement with the experimental data\cite{data}.
For comparison, we also include the results (in square brackets of the
second column of Table 1) of $f_{+}(0), \lambda_{+}, 
\lambda_{0}$, and $\xi_{A}$ obtained from 
the valence contribution in $q^{+}$$\neq$0 frame. 
Even though the form factor $f_{+}(0)$ in $q^{+}$$\neq$0
frame is free from the nonvalence contributions, its derivative at
$q^{2}$=0, i.e., $\lambda_{+}$, receives the nonvalence contributions. 
Moreover, the form factor $f_{-}(q^{2})$ in $q^{+}$$\neq$0 frame is not 
immune to the nonvalence contributions even at $q^{2}$=0\cite{Zm}.
Unless one includes the nonvalence contributions in the $q^{+}$$\neq$0
frame, one cannot really obtain reliable predictions for the observables 
such as $\lambda_{+}, \lambda_{0}$ and $\xi_{A}$ for $K_{l3}$ decays.
 
In Fig. 2, we show the form factors $f_{+}$ obtained from both $q^{+}$=0
and $q^{+}$$\neq$0 frames for $0\leq q^{2}\leq (M_{K}-M_{\pi})^{2}$ region.
As one can see in Fig. 2, the form factor $f_{+}(q^{2})$
obtained from $q^{+}$=0 frame (solid lines)\cite{Kaon} 
appears to be linear functions of $q^{2}$ justifying 
Eq. (5) usually employed in the analysis of experimental data\cite{data}. 
Note, however, that the $f_{+}(q^{2})$ obtained from only valence 
contribution in $q^{+}$$\neq$0 frame (dotted lines) does not exhibit the 
same behavior.
\begin{figure}
\centerline{\psfig{figure=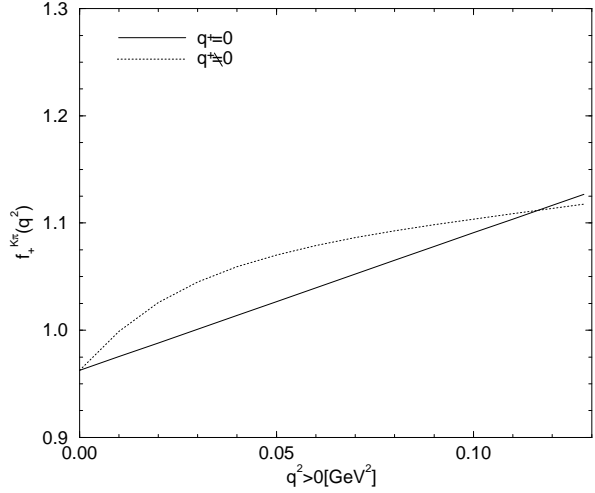,height=3 in}}
\caption{The form factors $f_{+}(q^{2})$ for the
$K\to\pi$ transition in timelike momentum transfer $q^{2}>0$. The solid
and dotted lines are the results from the $q^{+}$=0 and $q^{+}$$\neq$0
frames[2], respectively.
The differences of the results between the two frames are the measure of
the nonvalence contributions from $q^{+}$$\neq$0 frame.}
\end{figure}
\paragraph*{Heavy-to-light(heavy) decays:}
Our predicted decay rates for $D\to K$ and $D\to\pi$ are
$\Gamma(D^{0}\to K^{-}e^{+}\nu_{e})= 8.36 |V_{cs}|^{2}\times 10^{-2}$
ps$^{-1}$ and $\Gamma(D^{0}\to\pi^{-}e^{+}\nu_{e})=0.113 |V_{cd}|^{2}$
ps$^{-1}$, respectively. 
Using $|V_{cs}|=1.04\pm 0.16$ and $|V_{cd}|=0.224\pm 0.016$\cite{data}, 
we obtain the branching ratio of $Br(D\to K)= (3.75\pm 1.16)\%$
and $Br(D\to\pi)=(2.36\pm 0.34)\times 10^{-3}$, 
while the experimental data are $(3.66\pm 0.18)\%$ for $D\to K$ and 
$(3.9^{+2.3}_{-1.1}\pm 0.4)\times 10^{-3}$. Also, our predicted 
decay rates for $B\to\pi$ and $B\to D$ 
are $\Gamma(B^{0}\to\pi^{-}\ell^{+}\nu_{\ell})=8.16 |V_{ub}|^{2}$ ps$^{-1}$,
$\Gamma(B^{0}\to D^{-}\ell^{+}\nu_{\ell})= 9.39 |V_{cb}|^{2}$, 
respectively.  Using $|V_{ub}|=(3.3\pm 0.4\pm0.7)\times 10^{-3}$ and 
$|V_{bc}|=0.0395\pm0.003$\cite{data},
we obtain $Br(B\to\pi)=(1.40\pm 0.34)\times 10^{-4}$ and 
$Br(B\to D)= (2.28\pm 0.20)\%$. Our results
are quite comparable with the recent experimental data\cite{data},
$(1.8\pm0.6)\times 10^{-4}$ for $B\to\pi$ and 
$(2.00\pm 0.25)\%$ for $B\to D$, within the given error range.

In the heavy quark limit $M_{1(2)}$$\to$$\infty$, the form factor
$f_{+}(q^{2})$ is reduced to the universal Isgur-Wise (IW)
function, $\xi(v_{1}\cdot v_{2})= [2\sqrt{M_{1}M_{2}}/ (M_{1}+M_{2})]
f_{+}(q^{2})$, where $v_{1(2)}=P_{1(2)}/M_{1(2)}$.
Our prediction of the slope $\rho^{2}$=0.8 of the IW function at the
zero-recoil point defined as $\xi(v_{1}\cdot v_{2})=
1-\rho^{2}(v_{1}\cdot
v_{2}-1)$ is quite comparable with the current world average
$\rho_{\rm avg.}$=0.66$\pm$0.19\cite{data} extracted from exclusive
semileptonic $\bar{B}$$\to$$D\ell\bar{\nu}$ decay.

In conclusion, we analyzed the exclusive $0^{-}$$\to$$0^{-}$
semileptonic meson decays using the LFQM constrained by the
variational principle for the QCD-motivated effective Hamiltonian with 
the well-known linear plus Coulomb interaction\cite{Mix,PLB}.
The form factors $f_{\pm}$ are obtained in $q^{+}$=0 frame and then
analytically continued to the timelike region by changing $q_{\perp}$ to
$iq_{\perp}$ in the form factors. The matrix element of the ``${\perp}$"
component of the current $J^{\mu}$ is used to obtain the form factor
$f_{-}$. Our model provided overall a good agreement with the 
available experimental data and the lattice QCD results for the 
transition form factors and branching ratios of the $0^{-}\to 0^{-}$ 
semileptonic meson decays. Also it rendered a large
number of predictions to the heavy meson mass spectra and decay
constants\cite{PLB}. 
We think that the success of our model hinges on the advantage of
light-front quantization realized by the rational energy-momentum
dispersion relation. It is crucial to calculate the ``good" components
of the current in the reference frame which deletes the
complication from the nonvalence Z-graph contribution.
The present work broadens the utility of the standard light-front frame
\`{a} la Drell-Yan-West to the timelike form factor calculation.
We anticipate further stringent tests of our model with more accurate
data from future experiments and lattice QCD calculations.

\section*{Acknowledgement}
We thank Professor Dong-Pil Min for his hospitality during our stay 
at the Center for Theoretical Physics at Seoul National University.
This work was supported by the U.S. Department of
Energy(DE-FG-02-96ER4-0947). The North Carolina Supercomputing Center
and the National Energy Research Scientific Computing Center are also
acknowledged for the grant of computing time allocation.


\begin{references}
\bibitem{LB} S. D. Drell and T. M. Yan,
\Journal{\PRL}{24}{181}{1970}; G. West, \Journal{\PRL}{24}{1206}{1970};
G. P. Lepage and S. J. Brodsky, \Journal{\PRD}{22}{2157}{1980}.

\bibitem{Kaon} H.-M. Choi and C.-R. Ji, \Journal{\PRD}{59}{034001}{1999}.

\bibitem{CJC} P. L. Chung, F. Coester, and W. N. Polyzou,
\Journal{\PLB}{205}{545}{1988};
W. Jaus, \Journal{\PRD}{44}{2851}{1991};
F. Cardarelli et al., \Journal{\PLB}{332}{1}{1994};
\Journal{\PRD}{53}{6682}{1996}.

\bibitem{NPA} H.-M. Choi and C.-R. Ji, \Journal{\NPA}{618}{291}{1997};
{\it ibid}. {\bf 56}, 6010 (1997).

\bibitem{Mix} H.-M. Choi and C.-R. Ji, \Journal{\PRD}{59}{074015}{1999}.

\bibitem{Zm} H.-M. Choi and C.-R. Ji, \Journal{\PRD}{58}{071901}{1998}.

\bibitem{BH} S. J. Brodsky and D. S. Hwang, \Journal{NPB}{543}{239}{1998}.

\bibitem{Isgur} S. Godfrey and N. Isgur, \Journal{\PRD}{32}{189}{1985};
N. Isgur, D. Scora, B. Grinstein, and M. B. Wise,
\Journal{\PRD}{39}{799}{1989};
D. Scora and N. Isgur, \Journal{\PRD}{52}{2783}{1992}.

\bibitem{PLB}  H.-M. Choi and C.-R. Ji, ``Light-Front Quark Model
Analysis of Exclusive $0^{-}\to0^{-}$ Semileptonic Heavy Meosn Decays",
to be published in {\em Phys. Lett. B} [hep-ph/9903496].

\bibitem{Flynn} J. M. Flynn and C. T. Sachrajda, ``Heavy Quark
Physics From Lattice QCD", to appear in Heavy Flavor (2nd edition) edited by
A. J. Buras and M. Lindner (World Scientific, Singapore),
hep-lat/9710057.

\bibitem{Dem} N.B. Demchuk, I. L. Grach, I. M. Narodetskii, and
S. Simula, \Journal{\PAN}{59}{2152}{1996};
H.-Y. Cheng, C.-Y. Cheng, and C.-W. Hwang,
\Journal{\PRD}{55}{1559}{1997}.

\bibitem{data} Particle Data Group, C. Caso et al., 
\Journal{\EPJ}{3}{1}{1998}. 

\bibitem{Gatto} M. Ademollo and R. Gatto, \Journal{\PRL}{13}{264}{1964}.

\bibitem{Roos} H. Leutwyler and M. Roos, \Journal{\ZPC}{25}{91}{1984}.

\bibitem{Buck} A. Afanasev and W. W. Buck, \Journal{\PRD}{55}{4380}{1997}.

\bibitem{SI} D. Scora and N. Isgur, \Journal{\PRD}{52}{2783}{1995}.
\end{references}
\end{document}